\documentstyle[preprint,aps]{revtex}

\tightenlines

\begin{document}

\draft

\title{Addendum to: Radiative corrections to the Dalitz plot of semileptonic decays of neutral baryons with light or charm quarks}

\author{A. Mart\'{\i}nez$^a$, D.M. Tun$^b$, 
A. Garc\'{\i}a$^c$, and G. S\'anchez-Col\'on$^c$}

\address{$^a$Departamento de F\'{\i}sica. 
Escuela Superior de F\'{\i}sica y Matem\'aticas del IPN. \\
Unidad Profesional Adolfo L\'opez Mateos, Edificio 9. \\
Col. Lindavista, C.P. 07738, M\'exico, D.F., MEXICO.}

\address{$^b$Coordinaci\'on de Din\'amica Orbital. 
Telecomunicaciones de M\'exico. \\
Av. de las Telecomunicaciones s/n. 
C.P. 09300, M\'exico, D.F., MEXICO.}

\address{$^c$Departamento de F\'{\i}sica. 
Centro de Investigaci\'on y de Estudios Avanzados del IPN. \\
A.P. 14-740, C.P. 07000, M\'exico, D.F., MEXICO.}

\date{October 20, 1993}

\maketitle

\begin{abstract}
We show that the radiative corrections containing terms up to order 
$\alpha q / \pi M_1$ for unpolarized semileptonic decays of baryons with positron emission can be obtained by simply reversing the sign of the axial-vector form factors in the corresponding f\/inal expressions of such decays with electron emission. 
This rule is valid regardless of the f\/inal kinematical variables chosen and of the particular Lorentz frame in which the f\/inal results are required.
\end{abstract}


\paragraph*{I. Introduction.}

In the study of radiative corrections to semileptonic decays of ba-ryons~\cite{tres} one is forced, in order to perform the calculations involved, to choose the sign of the charge of the emitted charged lepton. 
Customarily, the negative sign is chosen. 
Nevertheless, the analysis and the procedures of such study are equally valid for the case when a positron is emitted. 
The rule to obtain the results for $e^+$ emission from the results of $e^-$ emission is to use the four momentum of the $\overline\nu$ as the $e^+$ momentum and the four momentum of the $e^-$ as the $\nu$ momentum. 
This is the same rule that applies to the non-radiatively corrected transition amplitude~\cite{uno}.

This rule can be applied directly to the f\/inal result even after it is already especialized to the rest frame of the decaying baryon provided the four momentum transfer is neglected in the radiative corrections, along with the $e^-$ mass, and still both the $\overline\nu$ and $e^-$ variables appear in the final results. 
Unfortunately, these conditions are not met by the Dalitz plot studied in Ref.~\cite{tres}.
One must make the above replacements in Ref.~\cite{tres} at a stage where both the $e^-$ and $\overline\nu$ four momenta are present and the result is still explicitly covariant. 
From this stage until the final result is especialized to the rest-frame of the decaying or emitted baryon a considerable amount of algebraic effort is still required.

It is the purpose of this note to introduce and demonstrate a very simple and practical rule that replaces the above one, that can be directly applied to the final not explicitly covariant $e^-$ results, and that does not require that either the $e^-$ or $\overline\nu$ variables appears in it.

This rule is: the radiatively corrected differential decay rate of semileptonic baryon decays with $e^+$ emission can be obtained from the one with $e^-$ emission by reversing the sign of the axial-vector form factors in this latter result, independently of the choice of the final kinematical variables and of the final reference frame.

In the next section we obtain this rule for the non-radiatively corrected differential decay rate. 
In Sect.~III we extend the rule when virtual radiative corrections are introduced and in Sect.~IV we incorporate bremsstrahlung radiative corrections to this rule. 
The last section contains our final discussions and conclusion. 
Throughout this paper we shall assume that no polarizations are observed and that radiative corrections are refined so as to include $\alpha q / \pi M_1$ contributions, where $q$ is the momentum transfer and $M_1$ is the mass of the decaying baryon. 
Also, in order to keep our analysis simple we shall assume time reversal invariance and, thus, that all the form factors involved can be chosen relatively real~\cite{dos}.

\paragraph*{II. The Uncorrected Differential Decay Rate.}

We now proceed to obtain the rule without considering the radiative corrections.
The uncorrected amplitude ${\cal M}_0$ for the processes 
$A^{{}^{0\atop -}} \rightarrow B^{{}^{+\atop 0}}\, e^-\, \overline{\nu}$ 
and
$A^{{}^{+\atop 0}} \rightarrow B^{{}^{0\atop -}}\, e^+\, \nu$
is respectively given by
$
{\cal M}_0 = (G_v/\sqrt 2)\, H_{\lambda}\, L^{{}^{(\mp)}}_{\lambda}
$
where $H_{\lambda}$ and $L^{{}^{(\mp)}}_{\lambda}$ are the hadronic and leptonic covariants. 
$H_{\lambda}$ is common to both groups of decays and is given by
$
H_{\lambda} = \overline{u}_B(p_2)\, {\cal W}_{\lambda}\, u_A(p_1)
$
where $p_1\,(p_2)$ is the four momentum of hyperon $A\,(B)$. 
${\cal W}_{\lambda}$ is the hadronic weak-interaction vertex, expressed as
$
{\cal W}_{\lambda} = 
\gamma_{\lambda}\,(f_1+g_1\gamma_5) +
\sigma_{\lambda\alpha}\,(q_{\alpha}/M_1)\,(f_2+g_2\gamma_5) +
(q_{\lambda}/M_1)\,(f_3+g_3\gamma_5) 
$

\noindent
with $f_i$ and $g_i$ $(i=1,2,3)$ the Dirac form factors and $q=p_1-p_2$ the four momentum transfer. For the decay with the $e^-$ emission we have 
$L^{{}^{(-)}}_{\lambda}={\overline u}_l\,O_{\lambda}\,v_{\nu}$,
and for the decay with the $e^+$ emission we have
$L^{{}^{(+)}}_{\lambda}={\overline u}_{\nu}\,O_{\lambda}\,v_l$.
Here 
$O_{\lambda}=\gamma_{\lambda}\,(1+\gamma_5)$
and our metric and $\gamma$-matrix conventions are those of Ref.~\cite{tres}.

After squaring ${\cal M}_0$ and summing over the spins, we obtain a result which is proportional to the product of the hadronic and leptonic traces,

\begin{equation}
\sum_s |{\cal M}_0|^2 = C\,
Tr[({\rlap/p}_2 + M_2)\,{\cal W}_{\lambda}\,
({\rlap/p}_1 + M_1)\,{\overline{\cal W}}_{\mu}]\,
Tr[{\rlap/l}\,\gamma_{\lambda}\,{\rlap/p}_{\nu}\,\gamma_{\mu}\,(1\pm\gamma_5)].
\label{cuatro}
\end{equation}
 
\noindent
$C$ is a constant whose explicit value is of no relevance here. 
The upper (lower) sign in Eq.~(\ref{cuatro}) corresponds to the decay with emitted $e^-$ $(e^+)$ and $M_1$ and $M_2$ are the masses of the baryons $A$ and $B$, respectively. 
In the leptonic traces $l$ and $p_{\nu}$ represent the four momenta of the electron or positron and the antineutrino or neutrino, respectively.

The hadronic trace above is a quadratic function of all the form factors.
That is, the hadronic trace will be given by a sum of the products of two form factors each one multiplied by certain traces of $\gamma$-matrices. 
These traces will either not contain a $\gamma_5$ or they will contain it only once. 
The, so to speak, pure products of form factors $f_i\/f_j$ and $g_i\/g_j$ will be accompanied by traces without a $\gamma_5$ and the cross products $f_i\/g_j$ will be accompanied by traces that contain the $\gamma_5$ only once. 
From standard theorems on the traces of $\gamma$-matrices we know that the trace of a product of $\gamma$-matrices none of which is a $\gamma_5$ is a real number and that the trace  of a product of $\gamma$-matrices only one of which is a $\gamma_5$ is an imaginary number. 
Therefore, the general form of the hadronic trace of Eq.~(\ref{cuatro}) will be a real part containing only pure products of form factors and an imaginary part containing only the cross products of form factors. 

The leptonic traces in Eq.~(\ref{cuatro}) can also be split into real and imaginary parts. One can readily see that their real parts are equal and that their imaginary parts have opposite sign, i.e., the leptonic traces above differ only by the sign of their imaginary parts.

The last step to obtain the rule is to notice that since the l.h.s. of Eq.~(\ref{cuatro}) must necessarily be a real number, its r.h.s. can be given only by the difference of the product of the real parts of the hadronic and leptonic traces and the product of their imaginary parts in the case with $e^-$ emission or by the sum of such products with $e^+$ emission. 
In other words, if $a+ib$ is the hadronic trace and $c+id$ and $c-id$ are the leptonic traces corresponding to $e^-$ and $e^+$ emission, 
then the r.h.s. of Eq.~(\ref{cuatro}) is necessarily given by $ac-bd$ and $ac+bd$, respectively.
Here $a$ is a function of $f_i\/f_j$ and $g_i\/g_j$ and $b$ is a function of $f_i\/g_j$ only. 

The rule then follows immediately. 
The expression of the differential decay rate for $e^+$ emission is obtained from the expression of the differential decay rate for $e^-$ emission by reversing the sign of the axial form factors $g_i$ in the latter~\cite{cuatro}.

\paragraph*{III. Virtual Radiative Corrections.}

The rule, $\lq\lq g_i \rightarrow -g_i"$, may be applied also in the case when the radiative corrections up to order $\alpha q/\pi M_1$ are included. 
We shall derive it by first considering the virtual part of these radiative corrections. 
In the next section the corresponding bremsstrahlung part will be considered.

For the sake of clarity we shall display the amplitudes with virtual radiative corrections. 
The notation and conventions are those of Ref.~\cite{tres}. 
For the decays 
$A^0 \rightarrow B^{{}^{\pm}}\, e^{{}^{\mp}}\, \nu$
they are given by

\begin{equation}
{\cal M}_{v_1}= 
\frac{G_v}{\sqrt 2} \frac{\alpha}{4 \pi^3 i}
\int \! d^4 k
(\frac{D_{\mu\alpha}}{k^2 - 2l\cdot k + i\epsilon}) 
\big[ \frac{(2p_{2\mu} + k_{\mu})H_{\lambda}}{k^2 + 2p_2\cdot k +i\epsilon} 
+ {\overline u}_B  T_{\mu\lambda}  u_A \big] 
{
{\overline u}_l
(2l_{\alpha} - \gamma_{\alpha}{\rlap/k})  O_{\lambda}  v_{\nu} 
\brace
{\overline u}_{\nu} O_{\lambda} 
(2l_{\alpha} - {\rlap/k}\gamma_{\alpha}) v_l 
}
\label{cinco}
\end{equation}

\begin{equation}
{\cal M}_{v_2} = 
\frac{G_v}{\sqrt 2} \, \frac{\alpha}{8 \pi^3 i} \, H_{\lambda}
\int \! d^4 k\, D_{\mu\alpha} \,
{ 
{\overline u}_l \,
\frac{
(2l_{\mu} - \gamma_{\mu}{\rlap/k}) \,
{\rlap/l} \,
(2l_{\alpha} - {\rlap/k}\gamma_{\alpha}) \,
({\rlap/l}+m)
}
{
2m^2(k^2 - 2l\cdot k + i\epsilon)^2
}
\, O_{\lambda} \, v_{\nu}
\brace
{\overline u}_{\nu} \, O_{\lambda} \,
\frac{
({\rlap/l}-m) \,
(2l_{\mu} + \gamma_{\mu}{\rlap/k}) \,
{\rlap/l}\,
(2l_{\alpha} + {\rlap/k}\gamma_{\alpha})
}
{
2m^2(k^2 + 2l\cdot k + i\epsilon)^2}
\, v_l }
\label{seis}
\end{equation}

\begin{equation}
{\cal M}_{v_3} = 
\frac{G_v}{\sqrt 2} \, \frac{\alpha}{8 \pi^3 i} \, 
H_{\lambda} \, L^{{}^{(\mp)}}_{\lambda}
\int \! d^4 k\, D_{\mu\alpha} \,
\frac{(2p_2-k)_{\mu}(2p_2-k)_{\alpha}}
{(k^2-2p_2\cdot k+i\epsilon)^2} + {\cal M}'_3.
\label{siete}
\end{equation}

\noindent
$H_{\lambda}$ and $L^{{}^{(\mp)}}_{\lambda}$ were def\/ined in Sect.~II. The upper (lower) lines within the brackets correspond to $e^-$ $(e^+)$ emission.
$T_{\mu\lambda}$ and ${\cal M}'_3$ in Eqs.~(\ref{cinco}) and (\ref{siete}) are model dependent parts.
One can readily see that the procedure to extract the model independent part and to absorb the model dependent part into the already existing form factors (explained in detail in Ref.~\cite{tres}) is carried on for $e^+$ emission in exact parallelism with the case of $e^-$ emission. 
After performing the virtual photon $k$-integration and adding the three amplitudes to the uncorrected one, one arrives at
$
{\cal M}_v = {\cal M}'_0 + {\cal M}^i_v
$
as in Ref.~\cite{tres}. 
The model dependence of the virtual radiative corrections is absorbed into the already existing form factors in ${\cal M}_0$,
through the definition of effective form factors (see Eq.~(18) of Ref.~\cite{tres}); this is indicated by putting a prime on it. 
The model independent virtual correction is given by

\begin{equation}
{\cal M}^i_v = \frac{\alpha}{\pi} \, \frac{G_v}{\sqrt 2} \, H_{\lambda} \,
\bigg[L^{{}^{(\mp)}}_{\lambda} \, \phi \pm 
{{\overline u}_l \, {\rlap/p}_2 \, O_{\lambda} \, v_{\nu}
\brace 
{\overline u}_{\nu} \, O_{\lambda} \, {\rlap/p}_2 \, v_l} \,
\phi' \bigg].
\label{nueve}
\end{equation}

\noindent
We do not require here  the explicit expressions of the functions $\phi$ and $\phi'$, they can be found in Eqs.~(7) and (8) of Ref.~\cite{tres}.
Althought $\phi$ and $\phi'$ are in general complex functions, only their real parts will contribute to the order $\alpha q/\pi M_1$ in the unpolarized case we are studying. 

For the group of decays $A^{{}^{\mp}} \rightarrow B^0\,e^{{}^{\mp}}\,\nu$ the only change comes in replacing $p_2$ by $-p_1$ in Eqs.~(\ref{cinco}), (\ref{seis}), and (\ref{siete}). 
The leptonic covariants in the equations of these decays appear exactly as they do in Eqs.~(\ref{cinco})--(\ref{siete}) and the result of the $k$-integrations leads to the same Eq.~(\ref{nueve}), with the corresponding functions $\phi$ and $\phi'$ (see Ref.~[5]). 
Therefore the general form of Eq.~(\ref{nueve}) applies equally well to this group of decays.

${\cal M}_v$ must be squared and summed over spins. To order $\alpha$ the result will be the interference of the two amplitudes added to Eq.~(\ref{cuatro}) (with primed form factors), namely,

\begin{eqnarray}
\sum_s |{\cal M}_v|^2 
&=& 
\sum_s |{\cal M}'_0|^2  
+ C'\,
Tr[({\rlap/p}_2 + M_2){\cal W}_{\lambda}\,
({\rlap/p}_1 + M_1){\overline{\cal W}}_{\mu}] 
\nonumber\\
&&
\times\big\{
(Re\,\phi)
Tr[\rlap/l\gamma_{\lambda}{\rlap/p}_{\nu}\gamma_{\mu}(1\pm\gamma_5)]  
+ m (Re\,\phi')
Tr[{\rlap/p}_2\gamma_{\lambda}{\rlap/p}_{\nu}\gamma_{\mu}(1\pm\gamma_5)]
\big\}. 
\label{diez}
\end{eqnarray} 

\noindent
Here $m$ is the mass of $e^{\mp}$ and as before the upper (lower) sign refers to  $e^-$ $(e^+)$ emission. 
$C'$ is an overall constant whose explicit value is irrelevant here.

One can now clearly see that the arguments of Sect.~II to establish the rule go through just as well for Eq.~(\ref{diez}). 
The l.h.s. of this equation must necessarily be a real number and thus its r.h.s. is given by the difference of the products of the real parts and the product of the imaginary parts of the hadronic and leptonic traces in the case of $e^-$ emission or by the sum of such products in the case of $e^+$ emission. Since the hadronic trace is the same one of Sect.~II, the two emissions differ only by a sign between the latter difference or sum, which can be associated with the $g'_i$ form factors~\cite{cuatro} and, thus, the rule follows: 
the results of virtual radiative corrections for $e^+$ emission can be obtained from the results of $e^-$ emission by just reversing the sign of the axial-vector form factors.

\paragraph*{IV. Bremsstrahlung Radiative Corrections.}

We shall now consider the bremsstrahlung radiative corrections. For the sake of clarity, we shall display the amplitude ${\cal M}_B$ of such corrections for the cases $A^0\rightarrow B^{{}^{\pm}}\,e^{{}^{\mp}}\,\nu\,\gamma$. 
Its detailed discussion for $e^-$ emission is given in Ref.~\cite{tres} and we shall keep the same notation and conventions. 
${\cal M}_B$ is composed of three summands,
$
{\cal M}_B = {\cal M}_1 + {\cal M}_2 + {\cal M}_3
$,
where

\begin{equation}
{\cal M}_1 = \pm\,\frac{eG_v}{\sqrt 2}\,\epsilon_{\mu}\,
\big(\frac{l_{\mu}}{l\cdot k}-\frac{p_{2_{\mu}}}{p_2\cdot k}\big)\,
H_{\lambda}\,L^{{}^{(\mp)}}_{\lambda}
\label{doce}
\end{equation}

\begin{equation}
{\cal M}_2 = \pm\,\frac{eG_v}{\sqrt 2}\,\epsilon_{\mu}\,
H_{\lambda}\,\frac{1}{2l\cdot k}\,
{{\overline u}_l\,\gamma_{\mu}\,{\rlap/k}\,O_{\lambda}\,v_{\nu}
\brace
{\overline u}_{\nu}\,O_{\lambda}\,{\rlap/k}\,\gamma_{\mu}\,v_l}
\label{trece}
\end{equation}

\begin{equation}
{\cal M}_3 = \pm\,\frac{eG_v}{\sqrt 2}\,\epsilon_{\mu}\,
{\overline u}_B\,T_{\mu\lambda}\,u_A\,L^{{}^{(\mp)}}_{\lambda}.
\label{catorce}
\end{equation}

\noindent
The hadronic and leptonic covariants $H_{\lambda}$ and $L^{{}^{(\mp)}}_{\lambda}$ are defined in Sect.~II and $e$ is the charge of $e^-$ (a negative number). 
The upper and lower overall signs and indeces and the upper and lower lines within the brackets correspond to $e^-$ and $e^+$ emissions, respectively. 
The tensor $T_{\mu\lambda}$ is given by

\begin{eqnarray}
T_{\mu\lambda} 
&=& 
\frac{1}{2p_2\cdot k}\,
\big[-\gamma_{\mu}\,{\rlap/k}-\frac{\kappa_2}{e_b}\,\sigma_{\mu\alpha}\,
k_{\alpha}\,({\rlap/p}_2+M_2)\big]\,{\cal W}_{\lambda}  
+\frac{1}{2p_1\cdot k}\,{\cal W}_{\lambda}\,\frac{\kappa_1}{e_b}\, 
({\rlap/p}_1+M_1)\,\sigma_{\mu\alpha}\,k_{\alpha} 
\nonumber \\
&& +\big(\frac{p_{2_{\mu}}k_{\rho}}{p_2\cdot k}-g_{\mu\rho}\big)\,
\bigg[\sigma_{\lambda\rho}\,
\big(\frac{f_2+g_2\,\gamma_5}{M_1}\big) +
g_{\lambda\rho}\,\big(\frac{f_3+g_3\,\gamma_5}{M_1}\big)\bigg].
\label{quince}
\end{eqnarray}

\noindent
Here $e_b$ is the charge of the charged baryon present in this group of decays. We remind the reader that $\epsilon_{\mu}$ is the polarization of the emitted real photon and $\kappa_1$ and $\kappa_2$ are the magnetic moments of hyperons $A$ and $B$, respectively.

Squaring ${\cal M}_B$ and summing over spins and photon polarization, one 
obtains

\begin{eqnarray}
&&
\sum_{s,\epsilon}
|{\cal M}_B|^2 
= C"Re\sum_{\epsilon} \epsilon_{\alpha}\epsilon_{\beta} 
\Bigg\{
Tr[({\rlap/p}_2 + M_2){\cal W}_{\lambda}
({\rlap/p}_1 + M_1){\overline{\cal W}}_{\mu}]  
\nonumber\\
&&
\times
\bigg[
\big(
\frac{l_{\alpha}l_{\beta}}{(l\cdot k)^2}
- \frac{p_{2_{\alpha}}l_{\beta} + l_{\alpha}p_{2_{\beta}}}
       {(l\cdot k)(p_2\cdot k)}
+ \frac{p_{2_{\alpha}}p_{2_{\beta}}}{(p_2\cdot k)^2}
\big) 
Tr[{\rlap/l}\gamma_{\lambda}{\rlap/p}_{\nu}\gamma_{\mu}(1\pm\gamma_5)]
\nonumber\\
&&
+ \frac{1}{(2l\cdot k)^2} 
Tr[{\rlap/p}_{\nu}\gamma_{\lambda}{\rlap/k}\gamma_{\alpha}{\rlap/l}
\gamma_{\beta}{\rlap/k}\gamma_{\mu}(1\pm\gamma_5)]
+ \frac{2}{2l\cdot k}
\big(\frac{l_{\alpha}}{l\cdot k} - \frac{p_{2_{\alpha}}}{p_2\cdot k}\big)
Tr[{\rlap/p}_{\nu}\gamma_{\lambda}{\rlap/l}\gamma_{\beta}
{\rlap/k}\gamma_{\mu}(1\pm\gamma_5)]
\bigg]
\nonumber\\[5pt]
&&
+ 2
Tr[({\rlap/p}_2 + M_2)T_{\alpha\lambda}
({\rlap/p}_1 + M_1){\overline{\cal W}}_{\mu}]
\nonumber\\
&&
\times
\bigg[
\big(\frac{l_{\beta}}{2l\cdot k} - \frac{p_{2_{\beta}}}{p_2\cdot k}\big)
Tr[{\rlap/l}\gamma_{\lambda}{\rlap/p}_{\nu}\gamma_{\mu}(1\pm\gamma_5)]
+\frac{1}{2l\cdot k}
Tr[{\rlap/p}_{\nu}\gamma_{\lambda}{\rlap/l}\gamma_{\beta}
{\rlap/k}\gamma_{\mu}(1\pm\gamma_5)]
\bigg]
\Bigg\}.
\label{dieciseis}
\end{eqnarray}

\noindent
$C"$ is an overall constant containing $\alpha$ and $G_v$ and other factors. The upper signs correspond to $e^-$ emission and the lower ones to $e^+$ emission, respectively. 
In Eq.~(\ref{dieciseis}) the first three terms correspond to the squares of ${\cal M}_1$ and ${\cal M}_2$ and to their interference, in this order, and the fourth and fifth terms correspond to the interferences of ${\cal M}_1$ and ${\cal M}_2$ with ${\cal M}_3$. 
The square of ${\cal M}_3$ does not appear because it contributes to orders higher than $\alpha q/\pi M_1$. 
Using some theorems on traces of $\gamma$-matrices we have reordered the leptonic traces of $e^+$ emission so that they resemble as much as posible their counterparts of $e^-$ emission.

The bremsstrahlung radiative corrections for the decays   
$A^{{}^{\mp}}\rightarrow B^0\,e^{{}^{\mp}}\,\nu\,\gamma$
lead to a similar result of Eq.~(\ref{dieciseis}) (see Eqs.~(18), (19), and (20) of Ref.~\cite{seis}).  
In particular, the leptonic traces that appear in the corresponding result are exactly the same ones of Eq.~(\ref{dieciseis}).

In addition to the hadronic trace that we encountered in sections II and III, the substitution of $T_{\mu\lambda}$ of Eq.~(\ref{quince}) into Eq.~(\ref{dieciseis}) lead to the appearance of several more complicated new traces, namely,
$
Tr[({\rlap/p}_2 + M_2)\,\gamma_{\alpha}\,{\rlap/ k}\,{\cal W}_{\lambda}\,
({\rlap/p}_1 + M_1)\,{\overline{\cal W}}_{\mu}],\   
$ 
$
Tr[({\rlap/p}_2 + M_2)\,\sigma_{\alpha\rho}\,k_{\rho}\,
({\rlap/p}_2 + M_2)\,{\cal W}_{\lambda}\,
({\rlap/p}_1 + M_1)\,{\overline{\cal W}}_{\mu}],\   
$ 
$
Tr[({\rlap/p}_2 + M_2)\,\sigma_{\lambda\alpha}\,
(f_2 + g_2\,\gamma_5)\,
({\rlap/p}_1 + M_1)\,{\overline{\cal W}}_{\mu}],\   
$
and  
$
Tr[({\rlap/p}_2 + M_2)\,p_{2_{\alpha}}\,k_{\lambda}\,
(f_3 + g_3\,\gamma_5)\,
({\rlap/p}_1 + M_1)\,{\overline{\cal W}}_{\mu}].\   
$
Just as for the hadronic trace of sections II and III, these last traces each will have a real and an imaginary part and in each case the real part will be a function of pure products $f_i\/f_j$ and $g_i\/g_j$ only and the imaginary part will be a function of cross products $f_i\/g_j$ only.

We can now clearly see that the $\lq\lq g_i \rightarrow -g_i"$ rule of sections II and III applies equally well with the bremsstrahlung radiative corrections. Inspection of Eq.~(\ref{dieciseis}) shows that the real parts of the leptonic traces are the same for $e^-$ and $e^+$ emissions and that their imaginary parts differ only by a minus sign. 
Since, the l.h.s. of Eq.~(\ref{dieciseis}) must necessarily be a real number, its r.h.s. must be given by the difference of the product of the real parts of hadronic and leptonic traces and the product of the imaginary parts of such traces in the case of 
$e^-$ emission and by the sum of these two kinds of products for $e^+$ emission. 
The sign difference between this difference and this sum can be systematically attached to the $g_i$ form factors. 
Therefore, we can conclude that the bremsstrahlung radiative corrections for baryon semileptonic decays with $e^+$ emission are obtained from the corresponding corrections with $e^-$ emission by simply reversing the sign of the axial-vector form factors.

\paragraph*{V. Discussion and Conclusion.}

In summary, we have demonstrated that the unpolarized differential decay rate of semileptonic baryon decays with positron emission including radiative corrections with up to order $\alpha q/\pi M_1$ contributions can be obtained directly from the corresponding decay rate with electron emission by simple reversing the sign of the axial-vector form factors.

This $\lq\lq g_i \rightarrow -g_i"$ rule remains valid independently of the choice of final kinematical variables and of the choice of the frame where the expressions loose their explicit covariance.

Also, our analysis is valid regardless of any particular $q^2$-dependence of the form factors involved. So, for example, if the form factors are expanded up to linear powers of $q^2$, the final expressions will involve products of, so to speak, slope parameters $\lambda^f_i$ and $\lambda^g_i$ and form factors $f_i(0)$ and $g_i(0)$ evaluated at $q^2=0$. 
Then, the sign of the slopes $\lambda^g_i$ must be reversed in the expressions for $e^-$ emission in order to obtain the expressions for $e^+$ emission.

Let us remark that the rule is not limited to electron and positron emission. Nowhere in our analysis have we required that the charged lepton mass be negligible.
Therefore, the rule is also valid when $\mu^-$ and $\mu^+$ are emitted.
Finally, let us point that the above rule applies not only to the results of Ref.~\cite{tres}, but also to previous results cited in this reference.


\begin{references}
\bibitem{tres} A. Mart\'{\i}nez and A. Garc\'{\i}a, D.M. Tun. Phys. Rev. 
{\bf D47}, 3984 (1993).
In this paper one can find references to previous results.
\bibitem{uno} V. Linke, Nucl. Phys. {\bf B12}, 669 (1969).
\bibitem{dos} This is not an essential assumption, but relaxing it does 
require a more detailed treatment that would appreciably lengthen this note. 
\bibitem{cuatro} One could equally well choose to reverse instead the sign of the $f_i$. But, since the presence of the $\gamma_5$ is what really matters and this matrix is always associated with the $g_i$, we prefer to apply the sign reversal to the $g_i$.
\bibitem{cinco} A. Garc\'{\i}a and S.R. Ju\'arez W. Phys. Rev. {\bf D22}, 1132 (1980); {\bf D22}, 2923 (E) (1980).
\bibitem{seis} D.M. Tun, S.R. Juarez W., and A.~Garc\'{\i}a, Phys. Rev. 
{\bf D44}, 3589 (1991). 
\end{references}
\end{document}